\documentclass[12pt]{article}
\usepackage{amsmath,amssymb}

\makeatletter
\@addtoreset{equation}{section}
\def\theequation{\thesection.\arabic{equation}}
\makeatother

\def\un#1{\relax\ifmmode\@@underline#1\else
        $\@@underline{\hbox{#1}}$\relax\fi}

\let\du=\du                     

\def\a{\alpha}
\def\b{\beta}

\def\d{\delta}

\def\g{\gamma}
\def\h{\eta}

\def\j{\psi}
\def\k{\kappa}
\def\l{\lambda}
\def\m{\mu}
\def\n{\nu}

\def\p{\pi}
\def\q{\theta}
\def\r{\rho}
\def\s{\sigma}

\def\G{\Gamma}
\def\J{\Psi}

\def\ve{\varepsilon}

\def\co{{\cal O}}

\def\bo{{\raise-.3ex\hbox{\large$\Box$}}}               
\def\pa{\partial}                                       
\def\TH{{\raise.2ex\hbox{$\displaystyle \bigodot$}\mskip-4.7mu \llap H \;}}
\def\face{{\raise.2ex\hbox{$\displaystyle \bigodot$}\mskip-2.2mu \llap {$\ddot
        \smile$}}}                                      

\def\Bar#1{\overline{#1}}                       
\def\leftrightarrowfill{$\mathsurround=0pt \mathord\leftarrow \mkern-6mu
        \cleaders\hbox{$\mkern-2mu \mathord- \mkern-2mu$}\hfill
        \mkern-6mu \mathord\rightarrow$}
\def\dvec#1{\vbox{\ialign{##\crcr
        \leftrightarrowfill\crcr\noalign{\kern-1pt\nointerlineskip}
        $\hfil\displaystyle{#1}\hfil$\crcr}}}           
\def\dt#1{{\buildrel {\hbox{\LARGE .}} \over {#1}}}     

\def\frac#1#2{{\textstyle{#1\over\vphantom2\smash{\raise.20ex
        \hbox{$\scriptstyle{#2}$}}}}}                   
\def\sfrac#1#2{{\vphantom1\smash{\lower.5ex\hbox{\small$#1$}}\over
        \vphantom1\smash{\raise.4ex\hbox{\small$#2$}}}} 
\def\bfrac#1#2{{\vphantom1\smash{\lower.5ex\hbox{$#1$}}\over
        \vphantom1\smash{\raise.3ex\hbox{$#2$}}}}       
\def\afrac#1#2{{\vphantom1\smash{\lower.5ex\hbox{$#1$}}\over#2}}    

\def\[{\lfloor{\hskip 0.35pt}\!\!\!\lceil}
\def\]{\rfloor{\hskip 0.35pt}\!\!\!\rceil}

\def\du#1#2{_{#1}{}^{#2}}
\def\ud#1#2{^{#1}{}_{#2}}
\def\dud#1#2#3{_{#1}{}^{#2}{}_{#3}}
\def\udu#1#2#3{^{#1}{}_{#2}{}^{#3}}

\def\fracm#1#2{\hbox{\large{${\frac{{#1}}{{#2}}}$}}}
\def\ha{{\fracmm12}}
\def\tr{{\rm tr}}

\def\un{\underline}
\def\fracmm#1#2{{{#1}\over{#2}}}

\def\low#1{{\raise -3pt\hbox{${\hskip 0.75pt}\!_{#1}$}}}

\def\Dot#1{\buildrel{_{_{\hskip 0.01in}\bullet}}\over{#1}}
\def\dt#1{\Dot{#1}}
\def\DDot#1{\buildrel{_{_{\hskip 0.01in}\bullet\bullet}}\over{#1}}
\def\ddt#1{\DDot{#1}}

\def\DDDot#1{\buildrel{_{_{\hskip 0.01in}\bullet\bullet\bullet}}\over{#1}}
\def\dddt#1{\DDDot{#1}}

\def\DDDDot#1{\buildrel{_{_{\hskip 
0.01in}\bullet\bullet\bullet\bullet}}\over{#1}}
\def\ddddt#1{\DDDDot{#1}}

\newskip\humongous \humongous=0pt plus 1000pt minus 1000pt

\newif\ifdtup

\newcommand{\be}{\begin{equation}}
\newcommand{\ee}{\end{equation}}
\newcommand{\nbe}{\begin{equation*}}
\newcommand{\nee}{\end{equation*}}

\newcommand{\fr}{\frac}
\newcommand{\lb}{\label}

\newcommand{\pd}{\partial}

\newcommand{\mr}{\mathrm}

\renewcommand{\(}{\left(}
\renewcommand{\)}{\right)}
\renewcommand{\[}{\left[}
\renewcommand{\]}{\right]}

\begin{document}

\thispagestyle{empty}

{\hbox to\hsize{
\vbox{\noindent September 2008 \hfill  revised version}}}

\noindent
\vskip2.0cm
\begin{center}

{\Large\bf ON THE SUPERSTRINGS--INDUCED \vglue.1in
FOUR-DIMENSIONAL GRAVITY, AND \vglue.2in 
ITS APPLICATIONS TO COSMOLOGY~\footnote{Supported in part 
by the Japanese Society for Promotion of Science (JSPS)}}

\vglue.3in

Masao Iihoshi~\footnote{Email address: iihosi-masao@ed.tmu.ac.jp}
and Sergei V. Ketov~\footnote{Email address: ketov@phys.metro-u.ac.jp}

\vglue.1in
{\it Department of Physics\\
     Tokyo Metropolitan University\\
     1--1 Minami-osawa, Hachioji-shi\\
     Tokyo 192--0397, Japan}
\end{center}
\vglue.1in
\begin{center}
{\Large\bf Abstract}
\end{center}
\vglue.1in
\noindent We review the status of the fourth-order (quartic in the spacetime 
curvature) terms induced by superstrings/M-theory (compactified on a warped 
torus) in the leading order with respect to the Regge slope parameter, and 
study their (non-perturbative) impact on the evolution of the Hubble scale in 
the context of the four-dimensional FRW cosmology. After taking into account 
the quantum ambiguities in the definition of the off-shell superstring 
effective action, we propose the generalized Friedmann equations, find the 
existence of their (de Sitter) exact inflationary solutions without a spacetime
 singularity, and constrain the ambiguities by demanding stability and the 
scale factor duality invariance  of our solutions. The most naive (Bel-Robinson
 tensor squared) quartic terms are ruled out, thus giving the evidence for the 
necessity of extra quartic (Ricci tensor-dependent) terms in the off-shell 
gravitational effective action for superstrings. Our methods are generalizable 
to the higher orders in the spacetime curvature. 

\newpage

\section{Introduction}

The homogeneity and isotropy of our Universe, as well as the observed spectrum
of density perturbations, are explained by inflationary cosmology \cite{inf}.
Inflation is usually realised by introducing a scalar field (inflaton) and 
choosing an appropriate scalar potential. When using Einstein equations, it
 gives rise to the massive violation of the strong energy condition and the
 exotic matter with large negative pressure. Despite of the apparent simplicity
 of such inflationary scenarios, the origin of their key ingredients, such as 
the inflaton and its scalar potential, remains obscure. 

Theory of superstrings is the leading candidate for a unified theory of 
Nature, and it is also the only known consistent theory of quantum gravity. It
is therefore natural to use superstrings or M-theory for the construction of
 specific mechanisms of inflation. Recently,  many brane inflation 
scenarios were proposed (see e.g. ref.~\cite{binf} for a review), together with
 their embeddings into the (warped) compactified superstring 
models, in a good package with the phenomenological constraints coming from 
particle physics (see e.g. ref.~\cite{kklt}). However, it did not contribute to
 revealing the orgin of the key ingredients of inflation. It also greatly 
increased the number of possibilities up to $10^{500}$ (known as the String 
Landscape), hampering specific theoretical predictions in the search for the 
signatures of strings and branes in the Universe. 

The inflaton driven by a scalar potential and their engineering by strings and 
branes are by no means required. Another possible approach can be based on a 
modification of the gravitational part of Einstein equations by terms of the 
higher order in the spacetime curvature \cite{starob}. It does not require an 
inflaton or an exotic matter, while the specific higher-curvature terms are 
well known to be present in the effective action of superstrings \cite{book}.

The perturbative strings are defined on-shell (in the form of quantum 
amplitudes), while they give rise to the infinitely many higher-curvature 
corrections to the Einstein equations, to all orders in the Regge slope 
parameter $\a'$ and the string coupling $g_s$. The finite form of all those
corrections is unknown and beyond our control. However, it still makes sense 
to consider the {\it leading} corrections to the Einstein equations, coming 
from strings and branes. Of course, any results to be obtained from the merely 
leading quantum corrections cannot be conclusive. Nevertheless, they may offer
 both qualitative and technical insights into the early Universe cosmology, 
within the well defined and highly restrictive framework.  
In this paper we adopt the approach based on the Einstein equations modified 
by the leading superstring-generated gravitational terms which are 
{\it quartic} in the spacetime curvature. We treat the quartic curvature terms
 {\it on equal footing} with the Einstein term, i.e. non-perturbatively.

We consider only geometrical (i.e. pure gravity) terms in the low-energy 
M-theory effective action in four space-time dimensions. We assume that the 
quantum $g_s$-corrections can be suppressed against the leading 
$\a'$-corrections, whereas all the moduli, including a dilaton and an axion, 
are somehow stabilized (e.g. by fluxes, after the warped compactification to 
four dimensions and spontaneous supersymmetry breaking).

Our paper is organized as follows. In Sec.~2 we review our starting point:  
M-theory in 11 spacetime dimensions with the leading quantum corrections, and 
the dimensional reduction to four spacetime dimensions. In Sec.~3 we discuss 
the problem of the off-shell extension of the gravitational part of the 
four-dimensional effective action for superstrings. In Sec.~4 we review the 
physical significance of the on-shell quartic curvature terms. In Sec.~5 we
prove  that it is impossible to eliminate the 4th order time derivatives in 
the 4-dimensional equations of motion with a generic metric. The structure of 
equations of motions for the special (FRW) metrics is revealed in Sec.~6, 
which contains our main new results. The exact (de Sitter) solutions, stability
 and duality constraints are also discussed in Sec.~6.  Our conclusion is 
Sec.~7. In Appendix A  we give our notation and compute some relevant 
identities. The two-component spinor formalism (for completeness) is summarized
 in Appendix B.

\section{M-theory and modified Einstein equations}

There are five perturbatively consistent superstring models in ten spacetime 
dimensions (see e.g. the book \cite{book}).  All those models are related by 
duality transformations. In this paper we are going to consider only 
the gravitational sector of the heterotic and type-II strings. In 
addition, there exists a parent theory behind all those superstring models, 
it is called M-theory, and it is eleven-dimensional \cite{book}. Not so much 
is known about the non-perturbative M-theory. Nevertheless, there are the
well-established facts that (i) the M-theory low-energy effective action is 
given by the 11-dimensional supergravity \cite{11dim}, and (ii) the leading 
quantum gravitational corrections to the 11-dimensional supergravity from 
M-theory in the bosonic sector are quartic in the curvature  \cite{mcor,mcor2}
 (see e.g. ref.~\cite{green} for some recent progress). Our purpose in this 
Section is to emphasize what is not known.

All the bosonic terms of the M-theory corrected 11-dimensional action read 
as follows \cite{mcor,mcor2}:
\begin{align}
S_{11} = & -\fracmm{1}{2\k^2_{11}}\int d^{11}x\,\sqrt{-g}\left[ R -
\fracmm{1}{2\cdot 4!}F^2 -\fracmm{1}{6\cdot 3!\cdot(4!)^2}\ve_{11}CFF\right]
\nonumber \\
& -\fracmm{T_2}{(2\p)^4\cdot 3^2\cdot 2^{13}}\int d^{11}x\,\sqrt{-g}
\left( J - \fracmm{1}{2} E_8 \right) + T_2 \int C\wedge X_8 \lb{11a}
\end{align}
where $\k_{11}$ is the 11-dimensional gravitational constant, $T_2$ is the 
M2-brane tension given by
\be T_2 = \left( \fracmm{2\p^2}{\k_{11}^2}\right)^{1/3}\lb{m2t} ~~,
\ee
$C$ is a 3-form gauge field of the 11-dimensional supergravity \cite{11dim}, 
and $F=d C$ is its four-form field strength, $R$ is the gravitational scalar 
curvature, $\ve_{11}$ stands for the 11-dimensional Levi-Civita symbol in the 
Chern-Simons-like coupling, while $(J,E_8,X_8)$ are certain {\it quartic\/} 
polynomials in the 11-dimensional curvature. The $J$ is given by 
\be J=3\cdot 2^8\left(R^{mijn}R_{pijq}R\du{m}{rsp}R\ud{q}{rsn}
+\fracmm{1}{2}R^{mnij}R_{pqij}R\du{m}{rsp}R\ud{q}{rsn}\right)
+{\cal O}(R_{mn})~,
\lb{oco}
\ee
the $E_8$ is the 11-dimensional extension of the eight-dimensional Euler 
density,
\be E_8= \fracmm{1}{3!}\ve^{abcm\low{1}n\low{1}\ldots m\low{4}n\low{4}}
\ve_{abc {m'}\low{1}{n'}\low{1}\ldots
{m'}\low{4}{n'}\low{4}}R\ud{{m'}\low{1}{n'}\low{1}}{m\low{1}n\low{1}}\cdots 
R\ud{{m'}\low{4}{n'}\low{4}}{m\low{4}n\low{4}} \lb{euler}
\ee
and the $X_8$ is the eight-form 
\be
X_8 =\fracmm{1}{192\cdot (2\p^2)^4}\left[ \tr R^4 - \fracmm{1}{4}(\tr R^2)^2
\right]~~, \lb{8form} \ee
where the traces are taken with respect to (implicit) Lorentz indices in 
eleven space-time dimensions. The (world) vector indices are also suppressed 
in eq.~(\ref{11a}).

The $J$-contribution (\ref{oco}) is defined {\it modulo\/} Ricci-dependent 
terms by its derivation \cite{mcor,mcor2}. The basic reason is the 
{\it on-shell\/} nature of the perturbative superstrings \cite{book}, whose 
quantum on-shell amplitudes determine the gravitational effective action modulo
 field redefinitions. Via the Einstein-Hilbert term, the metric field 
redefinitions contribute to the next (quartic) curvature terms with at least 
one factor of Ricci curvature.  Therefore, some additional physical 
requirements are needed in order to fix those Ricci-dependent terms in the 
off-shell M-theory effective action. 

To match the constraints imposed by particle physics, M-theory is supposed
to be compactified to one of the superstring models in ten dimensions, and then
down to four spacetime dimensions e.g., on a Calabi-Yau complex three-fold 
\cite{book}. Alternatively, M-theory may be directly compactified down to four
real dimensions on a 7-dimensional special $(G_2)$ holonomy manifold \cite{aw}.
The bosonic fields of the action (\ref{11a}) are just an eleven-dimensional 
metric and a 3-form (there is no dilaton in eleven dimensions). In other words,
the 11-dimensional action (\ref{11a}) is the most general starting point to 
discuss the M-theory/superstrings compactification. 

In the presence of fluxes, we should consider the {\it warped} 
compactification, whose metric is of the form \cite{dka}
\be 
ds^2_{11}=e^{2A(y)}ds^2_{\rm FRW} + e^{-2A(y)}ds^2_7~~,
\lb{warp}
\ee
where $ds^2_{\rm FRW}$ is the FRW metric in (uncompactified) four-dimensional
spacetime (see eq.~(\ref{frw1}) below), $ds^2_7$ is a metric in compactified 
seven dimensions with the coordinates $y^a$, $a=4,5,6,7,8,9,10$, and $A(y)$ is 
called  a warp factor.

Since we are interested in the gravitational sector of the 
four-dimensional type-II superstrings, an explicit form of the 7-metric 
$ds_7^2$ is not needed. In the case of heterotic strings, one has to  include 
the `anomalous' term quadratic in the curvature (see below). We put all the 
four-dimensional scalars (like a dilaton, an axion and moduli) into the matter 
stress-energy tensor (in Einstein frame), and assume that they are somehow 
stabilized to certain fixed values. In addition, we do not consider any 
M-theory/superstrings solitons such as M- or D-branes. After dimensional 
reduction, the only gravitational terms coming from type-II superstrings in 
four dimensions are given by
\be
S_4  =  -\fracmm{1}{2\k^2}\int d^{4}x\,\sqrt{-g}\left( R +\b J_R \right)
\lb{4a}
\ee
where we have introduced the Einstein coupling $\k$ in four dimensions, and
the four-dimensional counterpart $J_R$ of $J$ in eq.~(\ref{oco}), 
$i,j=0,1,2,3$, 
\be J_R= R^{mijn}R_{pijq}R\du{m}{rsp}R\ud{q}{rsn}
+\fracmm{1}{2}R^{mnij}R_{pqij}R\du{m}{rsp}R\ud{q}{rsn}+O(R_{mn})
\lb{oco4}
\ee

The relation between the coupling constants $\k_{11}$ and $\k$ is given by
\be 
\k^2=e^{5A}M_{\rm KK}^7\k_{11}^2 \lb{rco}
\ee
where we have introduced the {\it Kaluza-Klein} (KK) compactification scale 
$M_{\rm KK}^{-7}=Vol_7\equiv \int d^7y \sqrt{g_7}$ and the {\it average} warp 
factor $A$ (with an integer weight $p$),
\be 
e^{pA}= \fracmm{1}{Vol_7} \int d^7y \sqrt{g_7}\, e^{pA(y)}
\lb{awarp}
\ee
We also find
\be \b =\fracmm{1}{3}
\left( \fracmm{\k^2}{2^{23/2}\p^5 e^{14A}M_{\rm KK}^7}\right)^{2/3}
\lb{beta}
\ee
of mass dimension $-6$. For instance, when substituting the Planck scale 
$\k\approx 10^{-33}$cm and $M^{-1}_{\rm KK}\approx 10^{-15}$cm, and ignoring
the warp factor, $A=0$, we get the  incredibly small (and, in fact, 
unacceptable -- see Sec.~6) value
\be \b \approx 10^{-118}~{\rm cm}^{6} \lb{small}\ee 
 
As regards the four-dimensional heterotic strings, the action (\ref{4a}) is
to be supplemented by the term \cite{het2}
\be 
S_H  =  
-\fracmm{1}{2\k^2}\int d^{4}x\,\sqrt{-g}\left( \fracm{1}{8}J_H\right)
\lb{het}
\ee
where 
\be J_H = R_{ijkl}R^{ijkl} + {\cal O}(R_{mn}) \lb{oho}
 \ee
again modulo Ricci-dependent terms.
 
The gravitational action is to be added to a matter action, which lead to the
{\it modified\/} Einstein equations of motion (in the type II case, for 
definiteness)
\be
R_{ij}- \fracm{1}{2}g_{ij}R +\b \fracmm{1}{\sqrt{-g}}\fracmm{\d}{\d g^{ij}}
\left( \sqrt{-g}J_R\right) =\k^2 T_{ij} \lb{meo}
\ee
where $T_{ij}$ stands for the energy-momentum tensor of all the matter fields 
(including dilaton and axion).

Due to the ambiguities in the definition of the $J_R$-polynomial, it is also 
possible to replace it by
\be J_C= C^{mijn}C_{pijq}C\du{m}{rsp}C\ud{q}{rsn}
+\fracmm{1}{2}C^{mnij}C_{pqij}C\du{m}{rsp}C\ud{q}{rsn}+{\cal O}(R_{mn})
\lb{oco4c}
\ee
where we have introduced the Weyl tensor in four dimensions \cite{wald}, which 
is the traceless part of the curvature tensor -- See Appendices A and B.

\section{Going off-shell with the curvature terms}

There are about $10^2$ Ricci-dependent terms in the most general off-shell 
gravitational effective action that is {\it quartic} in the curvature. It 
also means about $100$ new coefficients, which makes the fixing of the 
off-shell action to be extremely difficult. The quartic curvature terms are 
thus different from the {\it quadratic} curvature terms, present in the 
on-shell  heterotic string effective action (\ref{het}), whose off-shell 
extension is very simple (see below). It is, therefore, desirable to formulate
 some necessary conditions that any off-shell extension has to satisfy.

(i) The first condition is, of course, the vanishing of all extra terms (i.e. 
beyond those in eq.~(\ref{oco4})) in the Ricci-flat case \cite{gwgz}. The 
perturbative superstring effective action is usually deducted from the 
superstring amplitudes, whose on-shell condition is just the Ricci-flatness. 
In the alternative method, known as the non-linear sigma-model beta-function 
approach, the Ricci-dependent ambiguities in the effective equations of motion
 (associated with the vanishing sigma-model beta-functions) arise via the 
dependence of the renormalization group beta-functions of the non-linear 
sigma-model upon the renormalization prescription, starting from two loops 
(see e.g. ref.~\cite{mybook} for details).

(ii) Supersymmetry requires all quantum bosonic corrections to be extendable to
 locally supersymmetric invariants. It can be made manifest in four spacetime 
dimensions, where the off-shell superspace formalism of $N=1$ supergravity is 
available \cite{wb}. The Weyl tensor, Ricci tensor and scalar curvature belong 
to three different $N=1$ superfields called $W_{\a\b\g}$, $G_{\a\dt{\a}}$ and 
$R$, respectively, while the first superfield is chiral.~\footnote{We use the 
two-component spinor notation \cite{wb}, $\a,\b,\ldots =1,2$ --- see Appendix 
B.} In particular, the Weyl tensor $C_{\a\b\g\d}$ appears in the first order 
of the $N=1$ superspace chiral anticommuting coordinates $\q^{\a}$ as 
\be  W_{\a\b\g}(x,\q) = W_{\a\b\g}(x)  + \q^{\d}C_{\a\b\g\d}(x) 
+ \ldots\lb{n1w}\ee
so that the $J_H$ terms (with all curvatures being replaced by Weyl tensors) 
is easily supersymmetrizable in superspace as
\be
\int d^2\q {\cal E}^{-1} W^2\low{\a\b\g}
\lb{sh}\ee
The $J_C$ terms in eq.~(\ref{oco4c}) are also extendable to the manifest 
superinvariant
\be \int d^4\q {E}^{-1} W^2\low{\a\b\g}\Bar{W}^2_{\dt{\a}\dt{\b}\dt{\g}}
\lb{sw4}\ee
where we have introduced the supervielbein densities ${\cal E}$ and $E$, 
in the chiral and central superspaces, respectively (see ref.~\cite{wb} for 
details). 

Those invariants were extensively studied in the past, because they naturally 
appear as the possible counterterms (with divergent coefficients) in quantum 
four-dimensional supergravity (see e.g. ref.~\cite{ulf}). In superstring theory
 one gets the {\it same\/} structures, though with {\it finite} coefficients 
(see e.g. refs.~\cite{free,stelle}). Thus, in four dimensions, the structure 
of the {\it on-shell superstrings} quartic curvature terms is fixed by local 
$N=1$ supersymmetry alone, up to normalization.

(iii) The absence of the higher order time derivatives is usually desirable to 
prevent possible unphysical solutions to the equations of motion, as well as 
preserve the perturbative unitarity, but it is by no means necessary. As is 
well known, the standard Friedmann equation of General Relativity is an 
evolution equation, i.e. it contains only the first-order time derivatives of 
the scale factor \cite{inf,ll}. It happens due to the cancellation of terms 
with the second-order time derivatives in the mixed $00$-component of Einstein 
tensor --- see e.g. Appendix of ref.~\cite{first} for details. It can also be 
seen as the consequence of the fact that the second-order dynamical 
(Raychaudhuri) equation for the scale factor in General Relativity can be 
integrated once, by the use of the continuity equation (\ref{em}), thus leading
 to the evolution (Friedmann) equation \cite{inf}. As regards the quadratic 
curvature terms present in the heterotic case, their unique off-shell extension
 is given by the Gauss-Bonnet-type combination \cite{bdes}
\be 
J_H\to G = R_{ijkl}R^{ijkl} - 4R_{ij}R^{ij} + R^2 \lb{gb}
\ee 
In the expansion around Minkowski space, 
$g_{ij}(x)=\h_{ij} + h_{ij}(x)$, the fourth-order derivatives (at the 
leading order in ${\cal O}(h^2)$) coming from the first term in eq.~(\ref{gb}) 
cancel against those in the second and third terms \cite{zwie}. As a 
result, the off-shell extension (\ref{gb}) appears to be ghost-free in all
dimensions. As regards {\it four} space-time dimensions, the terms (\ref{gb}) 
can be rewritten as the four-dimensional Euler density (\ref{eu}). Therefore, 
being a total derivative, eq.~(\ref{gb}) does not contribute to the 
four-dimensional effective action.~\footnote{Of course, adding Euler densities 
to the Einstein-Hilbert term matters in higher (than four) dimensions 
\cite{zumino,myers}, or with the dynamical dilaton and axion fields 
\cite{greek}.}

The higher time derivatives are apparent in the gravitational equations of 
motion with the quartic curvature terms (see also ref.~\cite{elizalde}). It is
 natural to exploit the freedom of the metric field redefinitions, in order 
to get rid of those terms. However, in Sec.~5 we prove that it is impossible 
to eliminate the 4th order time derivatives in the quartic curvature terms via
 a metric field redefinition. It may still be possible for some special 
(like FRW)  metrics, after imposing the string duality requirement (Sec.~6).  
  
(iv) The matter equations of motion in General Relativity imply the covariant 
conservation law of the matter energy-momentum tensor,
\be (T^{ij})_{;j}=0 \lb{em}
\ee
By the well known identity $(R^{ij}-\ha g^{ij}R)_{;j}=0$, eqs.~(\ref{meo}) and 
(\ref{em}) imply 
\be
\left[ \fracmm{1}{\sqrt{-g}}\fracmm{\d}{\d g^{ij}}
\left( \sqrt{-g}J\right)\right]_{;j}=0 \lb{div}
\ee
For instance, when $J=G$ as in eq.~(\ref{gb}), eq.~(\ref{div}) reads
\begin{align} 
-\ha (R_{ijkl}R^{ijkl} - 4R_{ij}R^{ij} + R^2)_{;m}+
2(R_{mjkl}R^{njkl})_{;n} \nonumber  \\
- 4(R_{minj}R^{ij})^{;n}-4(R_{mi}R^{in})_{;n}+2(RR_{mn})^{;n}  =0
\lb{div1}
\end{align}
By the use of Bianchi identities for the curvature tensor, we find by an
explicit calculation that the left-hand-side of eq.~(\ref{div1}) identically 
vanishes. We believe that eq.~(\ref{div}) should be {\it identically\/} 
satisfied by {\it any} off-shell gravitational correction $J$ because, 
otherwise, the consistency of the gravitational equations of motion may be
violated.

Given the quartic curvature terms (\ref{oco4}), the modified Einstein equations
 of motion (\ref{meo}) are 
\begin{align} 
\k^2 T_{ij} & = R_{ij}-\ha g_{ij}R +\b \left[  -\ha g_{ij}J_R 
 -R_{mhk(i}R\du{j)rt}{m} \left( R^{kqsr}R\udu{t}{qs}{h}+
R^{ksqt}R\ud{hr}{qs}\right)\right. \nonumber\\
& - R_{kqs(i}R_{j)rmt}\left( R^{hsqt}R\ud{krm}{k}-R^{thsq}R\du{h}{rmk}\right)
 + \left( R_{itrj}R^{ksqt}R\udu{h}{sq}{r}\right)_{(;k;h)}\lb{meo4}\\
&\left. + \left( R_{isqt}R^{rktm}R\dud{j}{sq}{k}\right)_{(;r;m)}
-\left( R\ud{hrs}{(i} R_{j)mnr}R\du{h}{mnk} + R\ud{sht}{(i} 
R_{j)mnl}R\ud{kmn}{h}\right)_{(;k;s)}
\right] \nonumber 
\end{align}

(v) We may also add the {\it causality\/} constraint as our next condition: 
the group velocity of ultra-violet perturbations on a gravitational background 
with the higher-curvature terms included, must not exceed the speed of light. 
As was demonstrated in ref.~\cite{gkny}, the causality condition merely affects
 the sign factors of the full curvature terms in the action, namely, the signs 
in front of $(R_{mnpq}R^{mnpq})^2$ and $(R_{mnpq}^*R^{mnpq})^2$ should be 
positive. It must be automatically satisfied by the perturbative superstring 
quartic corrections (\ref{oco4}) due to the known unitarity of superstring 
theory, and it is the case indeed ---  see the identity (\ref{imp}) --- 
just because $\b>0$.

Of course, our list is not complete, and it could be easily extended by more 
conditions, e.g. by requiring the consistency with black hole physics, 
gravitational waves, nucleosynthesis, etc. For example, in Sec.~6 we impose the
 scale factor duality as yet another constraint.

\section{On-shell structure and physical meaning of the quartic curvature 
terms}

The detailed structure and physical meaning of the quartic curvature terms in 
eqs.~(\ref{oco4}) and (\ref{oco4c}) are easily revealed via their connection to
 the four-dimensional {\it Bel-Robinson} (BR) tensor \cite{br}. The latter is 
well known in General Relativity \cite{penr,deser}. We review here the main 
properties of the BR tensor, and calculate the coefficients in the important 
identities -- see eqs.~(\ref{id1}) and (\ref{id2}) in this Section 
below.~\footnote{Those coefficients were left undetermined in 
ref.~\cite{deser}.}  

The BR tensor is defined by~\footnote{See also Appendix A for more.}
\be
T_R^{iklm}  = R^{ipql}R\udu{k}{pq}{m} +{}^*R^{ipql}{}^*R\udu{k}{pq}{m} 
\lb{belr}\ee
whose structure is quite similar to that of the Maxwell stress-energy tensor, 
\be T^{\rm Maxwell}_{ij}=F_{ik}F\du{j}{k}+{}^*F_{ik}{}^*F\du{j}{k}~~,\quad 
F_{ij}=\pa_iA_j-\pa_jA_i\lb{max}
\ee

The Weyl cousin $T_C^{ijlm}$ of the BR tensor is obtained by replacing all 
curvatures by 
Weyl tensors in eq.~(\ref{belr})--- see eq.~(\ref{wbrob}). The Weyl BR tensor 
can be 
factorized in the two-component formalism (see Appendix B),
\be (T_C)_{\a\b\g\d\dt{\a}\dt{\b}\dt{\g}\dt{\d}}=C\low{\a\b\g\d}
\bar{C}_{\dt{\a}\dt{\b}\dt{\g}\dt{\d}}
\lb{fwbr}\ee

In this section, we consider all the quartic terms on-shell, i.e. 
{\it modulo\/} 
Ricci-tensor dependent terms. Therefore, we are not going to distinguish 
between $T_R$ and 
$T_C$ here. The Ricci-tensor dependent additions will be discussed in 
Secs.~5 and 6.

The significance of the BR tensor to the quartic curvature terms is already 
obvious from 
superspace (see Sec.~3), where the locally N=1 supersymmetric extension of the 
quartic Weyl 
terms (\ref{oco4c}) is given by  eq.~(\ref{sw4}) whose bosonic part is the BR 
tensor 
squared, due to eq.~(\ref{fwbr}). As regards a straightforward proof, see 
Appendix A and 
our derivation of eq.~(\ref{imp}) there, which imply
\be
T^2_{ijkl}=8J_R = \frac{1}{4}(R_{ijkl}R^{ijkl})^2 
+\frac{1}{4}({}^*R_{ijkl}R^{ijkl})^2 \lb{id1}\ee
In addition, when using another identity (\ref{iden1}), eq.~(\ref{id1}) yields
\begin{align}
T^2_{ijkl}=8J_R = & -\frac{1}{4}({}^*R_{ijkl}^2)^2 
+\frac{1}{4}({}^*R_{ijkl}R^{ijkl})^2 
\nonumber\\
& = \frac{1}{4}(P_4^2-E^2_4) =\frac{1}{4}(P_4+E_4)(P_4-E_4) \lb{id2}\end{align}
where we have introduced the Euler and Pontryagin topological densities in 
four dimensions --- see eqs.~(\ref{eu}) and (\ref{pont}), respectively.

In addition \cite{br, deser}, the on-shell BR tensor is {\it fully symmetric} 
with respect to its vector indices, it is {\it traceless},
\be T_{ijkl}=T_{(ijkl)}~~,\qquad T^i_{ikl}=0~~, \lb{brs}\ee 
(ii) it is covariantly {\it conserved} (though the BR tensor is not a physical
 current!), 
\be \nabla^iT_{ijkl}=0~~, \lb{brcon}\ee
and it has {\it positive} `energy' density,
\be T_{0000}>0~~. \lb{brpos}\ee
Equation (\ref{brs}) is most easily seen in the two-component formalism 
(see Appendix B), eq.~(\ref{brcon}) is the consequence of Bianchi identities 
\cite{gare}, whereas eq.~(\ref{brpos}) just follows from the definition 
(\ref{belr}).

The BR tensor is related to the gravitational energy-momentum 
{\it pseudo}-tensors \cite{deser}. 
It can be most clearly seen in {\it Riemann Normal Coordinates} (RNC) at any 
{\it given}
 point in spacetime. The RNC are defined by the relations
\be  g_{ij}=\h_{ij}~,\quad g_{ij,k}=0~,\quad g_{ij,mn}=-\frac{1}{3}(R_{imjn}
+R_{injm})
\lb{rnci}\ee
so that the derivatives of Christoffel symbols read as follows: 
\be \G^{i}_{jk,l}=-\frac{1}{3}(R\ud{i}{jkl}+R\ud{i}{kjl})\lb{rncch}\ee
Raising and lowering of vector indices in RNC are performed with Minkowski 
metric $\h_{ij}$
 and its inverse $\h^{ij}$, whereas all traces in the last two 
eqs.~(\ref{rnci}) and 
(\ref{rncch}) vanish, 
\be \h^{ij}g_{ij,mn}=\h^{ij}\G^k_{ij,l}=\G^i_{ij,k}=\G^i_{jk,i}=0 
\lb{rctra}\ee

Moreover, there exists the remarkable non-covariant relation 
(valid only in RNC) \cite{deser}
\be T_{ijkl}=\pa_k\pa_l \left( t^{LL}_{ij}
+\frac{1}{2}t^E_{ij}\right)\lb{psetr}\ee 
where the symmetric {\it Landau-Lifshitz} (LL) gravitational pseudo-tensor 
\cite{ll}
\begin{align}
(t_{LL})^{ij} = &  -\h^{ip}\h^{jq}\G^k_{pm}\G^m_{qk} 
+\G^i_{mn}\G^j_{pq}\h^{mp}\h^{nq}
-\left( \G^m_{np}\G^j_{mq}\h^{in}\h^{pq}
+\G^m_{np}\G^i_{mq}\h^{jn}\h^{pq}\right) \nonumber \\
&  +h^{ij}\G^{m}_{np}\G^n_{mq}\h^{pq}\lb{pllt}
\end{align}
and the non-symmetric {\it Einstein} (E) gravitational  pseudo-tensor 
\cite{mtw} 
\be (t^E)^i_j = \left( -2\G^i_{mp}\G^m_{jq}+\d^i_j\G^n_{pm}\G^{m}_{qn}
\right)\h^{pq}
\lb{epst}\ee
have been introduced in RNC, in terms of Christoffel symbols.

\section{Off-shell quartic curvatures in cosmology}

The main Cosmological Principle of a {\it spatially\/} homogeneous and
isotropic $(1+3)$-dimensional universe (at large scales) gives rise to the
standard {\it Friedman-Robertson-Walker} (FRW) metrics of the form 
\cite{mtw}
\be
 ds_{\rm FRW}^2 = 
 dt^2 - a^2(t)\left[ \fracmm{dr^2}{1-kr^2} +r^2d\Omega^2\right]
\lb{frw1} \ee
where the function $a(t)$ is known as the scale factor in  `cosmic' 
coordinates 
$(t,r,\theta,\phi)$; we use $c=1$ and $d\Omega^2= d\theta^2 
+\sin^2\theta d\phi^2$, 
while $k$ is the FRW topology index taking values $(-1,0,+1)$. Accordingly, 
the FRW 
metric (\ref{frw1}) admits a 6-dimensional isometry group $G$ that is either 
$SO(1,3)$, $E(3)$ or $SO(4)$, acting on the orbits $G/SO(3)$, with  the spatial
3-dimensional sections $H^3$, $E^3$ or $S^3$, respectively. By the coordinate 
change, 
$dt=a(t)d\eta$, the FRW metric (\ref{frw1}) can be rewritten to the form
\be ds^2 = a^2(\eta)\left[ d\eta^2 - \fracmm{dr^2}{1-kr^2}
 - r^2d\Omega^2\right]
\lb{frw2} \ee 
which is manifestly (4-dim) conformally flat in the case of $k=0$. Therefore, 
the 4-dim Weyl tensor of the FRW metric obvioulsy vanishes in the `flat' case 
of $k=0$. It is well known that the FRW Weyl tensor vanishes in the other two 
cases, $k=-1$ and $k=+1$, too \cite{vweyl,first}. Thus we have 
\be C^{\rm FRW}_{ijkl}=0 \lb{frww}\ee

Inflation in an early universe is defined as the epoch during which the scale 
factor is accelerating \cite{inf},
\be \ddt{a}(t)>0~,~{\rm or~ equivalently}~~
 \fracmm{d}{dt}\left( \fracmm{H^{-1}}{a}\right)<0 
\lb{idef}
\ee
where the dots denote time derivatives, and $H=\dt{a}/a$ is Hubble `constant'.
 The amount of inflation is given by a number of e-foldings \cite{inf},
\be N =\ln \fracmm{a(t_{\rm end})}{a(t_{\rm start})}= 
\int^{t_{\rm end}}_{t_{\rm start}}
 H~dt \lb{efol}\ee 
which should be around $70$ \cite{inf}. 

Though the leading purely geometrical (perturbative) correction in the 
heterotic string case is given by the Gauss-Bonnet combination (\ref{gb}), and 
thus it does not contribute to the equations of motion in four space-time
dimensions, the situation changes when the {\it dynamical} moduli (axion and 
dilaton) are included. The effective string theory couplings are 
moduli-dependent, which gives rise to a non-trivial coupling with the moduli 
in front of the Gauss-Bonnet term, so that the latter  is not a total 
derivative any more. At the level of the one-loop corrected heterotic 
superstring effective action in four dimensions, the cosmological solutions 
were studied in ref.~\cite{greek}. As regards the realization of inflation in 
M-theory, see e.g. ref.~\cite{bbk}.

In the case of type-II superstrings (after stabilizing the moduli) we are left
 with the quartic curvature terms in the four-dimensional effective action
 (Sec.~2). Let's address the issue of the higher time derivatives in the 
general setting. It is quite natural to use the freedom of the metric field 
redefinitions in string theory in order to try to get rid of the higher time 
derivatives in the effective action. The successful example is provided by the
 Gauss-Bonnet gravity (Sec.~3) that we are now going to follow. Let's consider 
a weak gravitational field~\footnote{We assign the lower case {\it latin} 
letters to spacetime indices, $i,j,k,\ldots=0,1,2,3$, and the lower case
{\it middle greek} letters to spatial indices, $\m,\n,\ldots=1,2,3$.}
\be g_{ij}(x)=\h_{ij} + h_{ij}(x) \lb{weak}\ee
in the harmonic gauge
\be (h_{ij})^{,j} =\frac{1}{2}\pa_{i}h~,\quad  h=\h^{ij}h_{ij} 
\lb{harg}\ee
The linearized curvatures are given by
\be R_{ijkl}=\ha\left[ h_{il,jk}-h_{jl,ik}-h_{ik,jl}
+h_{jk,il}\right]
\lb{linc}\ee
whereas the Ricci tensor and the scalar curvature in the gauge (\ref{harg})
read
\be R_{ij} = -\ha \bo h_{ij}~,\quad R=-\ha \bo h~,\quad \bo \equiv \pa^{i}
\pa_{i} \lb{linri}\ee
As is clear from the structure of those equations, it is possible to form the 
Ricci terms after integration by parts in the quadratic curvature action. As a 
result, there is a cancellation of all terms with the 4th order time 
derivatives in the leading order ${\cal O}(h^2)$ of the Gauss-Bonnet action 
(\ref{gb}) in all spacetime dimensions, as was first observed in 
ref.~\cite{zwie}.

Unfortunately, we find that it does not work for the quartic curvature terms,
 even in four spacetime dimensions, as we now going to argue.

When using the linearized curvature (\ref{linc}), the quartic terms 
(\ref{oco4}) in four spacetime dimensions have the structure
\begin{align}
2^5 J_R
&= A^{ikjl}A_{iljk} + 2A^{ikjl}B_{iljk} + B^{ikjl}B_{iljk}
\notag \\                 
&\hspace{0.6cm}+ A^{ikjl} \{ C_{ilkj} + C_{lijk} \} 
                    + B^{ikjl} \{ C_{ilkj} + C_{lijk} \}
\notag \\ 
 &\hspace{0.6cm} + 2C^{ikjl} \{ C_{ilkj} + C_{lijk} + C_{kjil} + C_{jkli} \} 
\notag \\
 &\hspace{0.6cm} - C^{ikjl} \{ C_{iljk} + C_{likj} + C_{jkil} + C_{kjli} \}
\lb{linqu}
\end{align}                
where we have introduced the notation $\pa^2_{ij}=\pa_{i}\pa_{j}$ and
\begin{align}
A^{ikjl}
&= \pd^{2}_{mn}h^{ik}(\pd^{2mn}h^{jl} + \pd^{2jl}h^{mn}
- \pd^{2jm}h^{ln} - \pd^{2ln}h^{jm}), \nonumber\\
B^{ikjl} &= \pd^{2ik}h_{mn}(\pd^{2mn}h^{jl} + 
\pd^{2jl}h^{mn} - \pd^{2jm}h^{ln} - \pd^{2ln}h^{jm}), \\
C^{ikjl} &= \pd^{2i}_{m}h^{k}_{n}(\pd^{2mn}h^{jl} 
+ \pd^{2jl}h^{mn}  - \pd^{2jm}h^{ln} - \pd^{2ln}h^{jm}) 
\nonumber\lb{abcnot}
\end{align}
while all the index contractions above are performed with Minkowski metric.

Equation (\ref{linqu}) is not very illuminating, but it is enough to 
observe that the dangerous terms $(\pa^2_{00}h_{\m\n})^4$  and 
$(\pa_0\pa_{\l}h_{\m\n})(\pa^2_{00}h_{\m\n})^3$ do contribute, and thus lead 
to the terms with the 4th and 3rd order time derivatives in the equations of 
motion, when all $h_{\m\n}$ are supposed to be independent. The last 
possibility is to 
convert those terms into some Ricci-tensor dependent contributions. However, 
in the harmonic gauge (\ref{harg}), getting the Ricci tensor requires the two
 spacetime derivatives to be contracted into the wave operator, as in 
eq.~(\ref{linri}), in each dangerous term, which is impossible for the quartic 
curvature terms, unlike their quadratic counterpart, because any integration 
by parts in the quartic terms does not end up with a wave operator in each 
term. The equations of motion in the case of $(BR)^2$-gravity with 
the FRW metric are explicitly computed in the next Sec.~6, as an example.

Having failed to remove the higher time derivatives for a generic metric, one 
can try to get rid of them for a special class of metrics, namely, the FRW 
metrics of our interest. The simplest example arises when all the Riemann 
curvatures in the quartic curvature terms are replaced by the Weyl tensors, as 
in eq.~(\ref{oco4c}). It also amounts to adding certain quartic curvature terms
 with at least one Ricci factor to the effective action (\ref{4a}). This 
proposal is based on the reasonable assumption \cite{kall} coming from the 
AdS/CFT correspondence that the $AdS_7\times S^4$ and $AdS_4\times S^7$ spaces 
seem to be the exact solutions to the (eleven-dimensional) M-theory equations 
of motion. Of course, such assumption is just the sufficient condition, not the
 necessary one, because there may be many more solutions. The substitution  
$R_{ijkl}\to C_{ijkl}$ leads to the contributions with three Weyl 
tensors (from the quartic terms) in the equations of motion, which implies 
{\it no} perturbative superstring corrections to the FRW metrics at all, 
because of eq.~(\ref{frww}).

In the next Sec.~6 we find that the scale factor duality requirements allow a 
family of the generalized Friedmann equations coming from the most general 
quartic curvature terms, with just a few real parameters. 

\section{Exact solutions, stability and duality}

Our motivation in this paper is based on the observation that the Standard 
Model (SM) of elementary particles does not have an inflaton.~\footnote{The 
proposal \cite{schap} to identify the inflaton with the SM Higgs boson requires
 its non-minimal coupling to gravity, which does not fit to string theory.} In 
addition, M-theory/superstrings have plenty of inflaton candidates but any 
inflationary mechanism based on a scalar field is highly model-dependent. When 
one wants the universal geometrical mechanism of inflation based on gravity 
only, it should occur due to some Planck scale physics to be described by the 
higher curvature terms ({\sf cf.} ref.~\cite{starob}).

On the experimental side, it is known that the vacuum energy density 
$\r_{\rm inf}$ during inflation is bounded from above by a (non)observation of 
tensor fluctuations of the Cosmic Microwave Background (CMB) radiation 
\cite{cmbr}, 
\be \r_{\rm inf} \leq \left( 10^{-3}M_{\rm Pl}\right)^4 \ee     
It severly constrains but does not exclude the possibility of the geometrical 
inflation originating from the purely gravitational sector of string theory, 
because the factor of $10^{-3}$ above may be just due to some numerical 
coefficients ({\sf cf.} Sec.~2).

In this Section we consider the structure of our generalized Friedmann equation
 with {\it generic} quartic curvature terms. We get the conditions of stability
 of our inflationary solutions, and solve the duality invariance constraints 
coming from string theory \cite{sdual}.

Due to a single arbitrary function $a(t)$ in the FRW Ansatz (\ref{frw1}), it is
 enough to take only one gravitational equation of motion in eq.~(\ref{meo}) 
without matter, namely, its mixed $00$-component. As is well known \cite{inf}, 
the spatial (3-dimensional) curvature can be ignored in a very early universe, 
so we choose the manifestly conformally-flat FRW metric (\ref{frw1}) with 
$k=0$ in our Ansatz.  It leads to a purely gravitational equation of motion 
having the form 
\be \lb{4eqm} 3H^2\equiv
3\left(\fracmm{\dt{a}}{a}\right)^2=\b P_8\left( \fracmm{\dt{a}}{a},
\fracmm{\ddt{a}}{a}, \fracmm{\dddt{a}}{a},\fracmm{\ddddt{a}}{a}\right)~,
\ee
where $P_8$ is a {\it polynomial\/} with respect to its arguments, 
\be \lb{poly}
P_8= \sum_{n_1+2n_2+3n_3+4n_4=8,\atop n_1,n_2,n_3,n_4\geq 0} 
c_{n_1n_2n_3n_4} \left(\fracmm{\dt{a}}{a}\right)^{n_1}
\left(\fracmm{\ddt{a}}{a}\right)^{n_2}\left(\fracmm{\dddt{a}}{a}\right)^{n_3}
\left(\fracmm{\ddddt{a}}{a}\right)^{n_4} 
\ee
Here the sum goes over the {\it integer\/} partitions $(n_1,2n_2,3n_3,4n_4)$ of
 $8$, the dots stand for the derivatives with respect to time $t$, and 
$c_{n_1n_2n_3n_4}$ are some real coefficients. The highest derivative enters
linearly at most, $n_4=0,1$.

The FRW Ansatz with $k=0$ gives the following non-vanishing curvatures:
\be
R\ud{0}{\m 0\n}=\d_{\m\n}\ddt{a}a,~~ 
R\ud{\m}{\n\l\r}=\left(\d^{\m}_{\l}\d_{\n\r}-\d^{\m}_{\r}\d_{\n\l}\right)
(\dt{a})^2,~~
R^{\m}_{\n}=-\d^{\m}_{\n}
\left[ \fracmm{\ddt{a}}{a}+2 \left(\fracmm{\dt{a}}{a}\right)^2\right]
\lb{frwcurv}
\ee
where $\m,\n,\l,\r=1,2,3$.  For example, in the case of the
$(BR)^2$ gravity (\ref{meo4}), after a straightforward (though quite tedious) 
calculation of the mixed $00$-equation without matter and with the curvatures 
(\ref{frwcurv}), we find  
\begin{align}
3H^2 & + \beta\left[ 9\left( \fracmm{\ddt{a}}{a}\right)^4 -36 
H^2\left(\fracmm{\ddt{a}}{a}\right)^3 +84 H^4\left(\fracmm{\ddt{a}}{a}\right)^2
 -36H\left(\fracmm{\ddt{a}}{a}\right)^2
\left(\fracmm{\dddt{a}}{a}\right) \right.\nonumber\\
& \left. +63H^8  -72H^3\left(\fracmm{\ddt{a}}{a}\right)
\left(\fracmm{\dddt{a}}{a}
\right) + 48H^6\left(\fracmm{\ddt{a}}{a}\right)-24H^5
\left(\fracmm{\dddt{a}}{a}\right)\right]=0
\lb{br2frw}
\end{align}
It is remarkable that the 4th order time derivatives (present in various terms
 of eq.~(\ref{meo4})) cancel, whereas the square of the 3rd order time
derivative of the scale factor, $\dddt{a}{}^2$, does not appear at all in this 
equation.~\footnote{Taking Weyl tensors instead of Riemann curvatures leads to 
the vanishing coefficients.}

Our generalized Friedmann equation (\ref{4eqm}) applies to {\it any\/} 
combination of the quartic curvature terms in the action, including the 
Ricci-dependent terms. The coefficients $c_{n_1n_2n_3n_4}$ in eq.~(\ref{poly}) 
can be thought of as linear combinations of the coefficients in the most 
general quartic curvature action. The polynomial (6.3) merely has $12$
undetermined coefficients, that is considerably less than a $100$ of the 
coefficients in the most general quartic curvature action. 

The structure of eqs.~(\ref{4eqm}) and (\ref{poly}) admits the existence of 
rather generic exact inflationary solutions without a spacetime singularity. 
Indeed, when using the most naive (de Sitter) Ansatz for the scale  factor,  
\be  \lb{expf} a(t) =a_0 e^{Bt}  \ee
with some real positive constants $a_0$ and $B$, and substituting 
eq.~(\ref{expf}) into 
eq.~({\ref{4eqm}), we get $3B^2=(\#)\b B^8$, whose coefficient $(\#)$ is just 
a sum of all $c$-coefficients in eq.~(\ref{poly}). Assuming the $(\#)$ to be 
positive, we find an exact solution,
\be  \lb{simple}  B =  \left( \fracmm{3}{\#\b}\right)^{1/6} \ee
This solution in non-perturbative in $\beta$, i.e. it is impossible to get it 
when considering the quartic curvature terms as a perturbation. Of course, the 
assumption that we are dealing with the leading correction, implies 
$Bt\ll 1$. Because of eqs.~(\ref{beta}) and (\ref{simple}), it leads to the 
natural hierarchy
\be \k M_{\rm KK}\ll 1 \quad {\rm or}\quad  l_{\rm Pl}\ll l_{\rm KK}\ee
where we have introduced the four-dimensional Planck scale $l_{\rm Pl}=\k$ and 
the compactification scale $l_{\rm KK}=M_{\rm KK}^{-1}$.

The {\it effective} Hubble scale $B$ of eq.~(\ref{simple}) should be 
lower than the {\it effective} (with warping) KK scale
 $M^{\rm eff.}_{\rm KK}=e^A M_{\rm KK}$, in order to validate our 
four-dimensional description of gravity, i.e. the ignorance of all KK modes,
\be B < M^{\rm eff.}_{\rm KK} \label{effneq}
\ee
It rules out the naive KK reduction (with $A=0$) but still allows the warped 
compactification (\ref{warp}), when the average warp factor is tuned,
\be
e^A < \fracmm{(\k M_{\rm KK})^{2/5}}{(9/\#)^{3/10}2^{23/10}\p}
\sim {\co}\left(10^{-3}\right)\ee  
where we have used eq.~(\ref{beta}) and have estimated $(\#)$ by order $10$.

The exact solution (\ref{expf}) is non-singular, while it describes an 
inflationary isotropic and homogeneous early universe.~\footnote{The exact de 
Sitter solutions in the special case (\ref{oco4}) were also found in 
ref.~\cite{ohta}.} Given the expanding universe, the curvatures decrease, so 
that the higher curvature terms cease to be the dominant contributions against 
the matter terms we ignored in the equations of motion. The matter terms may 
provide a mechanism for ending the geometrical inflation and reheating (i.e. 
a Graceful Exit to the standard cosmology).

To be truly inflationary solutions, eqs.~(\ref{expf}) and (\ref{simple}) should
correspond to the stable fixed points (or attractors) \cite{inf}. The stability
 conditions are easily derived along the standard lines (see e.g.,  
refs.~\cite{sami,borunda}). When using the parametrization 
\be a(t) =e^{\l(t)}~, \lb{param} \ee
we easily find
\begin{align}
\fracmm{\dt{a}}{a}~=~& \dt{\l}~~, \\
\fracmm{\ddt{a}}{a}= ~& \ddt{\l} +(\dt{\l})^2~~,\nonumber \\
\fracmm{\dddt{a}}{a}~=~& \dddt{\l} + 3\ddt{\l}\dt{\l} +(\dt{\l})^3~~,
\nonumber \\
\fracmm{\ddddt{a}}{a}~=~& \ddddt{\l} +4\dddt{\l}\dt{\l}+
 6\ddt{\l}(\dt{\l})^2 + 3(\ddt{\l})^2+(\dt{\l})^4\nonumber 
\lb{lamb} 
\end{align}

Equation (\ref{poly}) now takes the form
\be \lb{poly2}
P_8= \sum_{n_1+2n_2+3n_3+4n_4=8,\atop n_1,n_2,n_3,n_4\geq 0} 
d_{n_1n_2n_3n_4} \left(\dt{\l}\right)^{n_1}
\left(\ddt{\l}\right)^{n_2}\left(\dddt{\l}\right)^{n_3}
\left(\ddddt{\l}\right)^{n_4} 
\ee
where the $d$-coefficients are linear combinations of the $c$-coefficients  
(easy to find). Equations (\ref{expf}) and (\ref{simple}) are also simplified,
\be \lb{simpler} 
\l(t) = Bt +\l_0~,\qquad {\rm where}\qquad  a_0=e^{\l_0}\quad {\rm and}\quad
  d_{8000}=\# ~.
\ee
The solution (\ref{simpler}) can be considered as the fixed point of the 
equations of motion (\ref{4eqm}) in a generic case,
\be
3y_1^2=\b P_8(y_1,y_2,y_3,\dt{y}_3)\equiv \b P_{8,0}(y_1,y_2,y_3) +
\b P_{4}(y_1,y_2,y_3)\dt{y}_3~,\lb{4eqm1}
\ee
where we have introduced the notation
\be  \lb{nota}
y_1=\dt{\l}~,\quad y_2=\ddt{\l}~,\quad y_3=\dddt{\l}~~. 
\ee
Equation (\ref{4eqm1}) can be brought into an autonomous form, 
\begin{align} 
\dt{y}_1~=~& y_2~~,\nonumber\\
\dt{y}_2~=~& y_3~~,\nonumber\\
\dt{y}_3~=~& \fracmm{3y_1^2 -\b P_{8,0}(y_1,y_2,y_3)}{\b P_{4}(y_1,y_2,y_3)}
\equiv f(y_1,y_2,y_3)  \lb{autom}
\end{align}
that is quite suitable for the stability analysis against small perturbations
about the fixed points, $y_a=y_a^{\rm fixed} + \d y_a$, where $a=1,2,3$. We 
find 
\begin{align} 
\d \dt{y}_1~=~& \d y_2~~,\nonumber\\
\d\dt{y}_2~=~& \d y_3~~,\nonumber\\
\d\dt{y}_3~=~& \left.\fracmm{\pa f}{\pa y_1}\right|\d y_1 
+ \left.\fracmm{\pa f}{\pa y_2}\right| \d y_2  
+  \left.\fracmm{\pa f}{\pa y_3}\right| \d y_3~~, \nonumber \\ 
\lb{perturb} 
\end{align}
where all the partial derivatives are taken at the fixed point (denoted by
$\left.{}\right|$). The fixed points are stable when all the eigenvalues of the
matrix
\be \lb{matrix}
\hat{M} = \left( \begin{array}{ccc}
 0 & 1 & 0 \\
0 & 0 & 1 \\
\left.\fracmm{\pa f}{\pa y_1}\right| & 
\left.\fracmm{\pa f}{\pa y_2}\right| & \left.\fracmm{\pa f}{\pa y_3}\right|
\end{array}
\right)
\ee
in eq.~(\ref{perturb}) are {\it negative} or have negative real parts
\cite{sami,borunda}. Then the fixed point is a stable attractor.

To the end of this Section, we would like to investigate how the symmetries of 
string theory are going to affect the coefficients of our generalized Friedmann
equation. Here we apply the scale factor duality \cite{sdual} by requiring our 
equation (\ref{4eqm}) to be invariant under the duality transformation
\be 
a(t) \leftrightarrow \fracmm{1}{a(t)} \equiv b(t) \lb{duality}
\ee
This duality is a cosmological version of the genuine stringy T-duality (which
is the symmetry of the non-perturbative string spectrum), in the case of 
time-dependent backgrounds. The scale factor duality is merely the symmetry of 
the (perturbative) equations of motion of the background fields. It is used 
e.g. in the so-called pre-big bang scenario \cite{vene}, in order to avoid the
cosmological singularity.

In the $\l$-parametrization (\ref{param}) the duality transformation
(\ref{duality}) takes the very simple form
\be 
\l(t) \leftrightarrow -\l(t) \lb{duality2}
\ee
The equations of motion in the form (\ref{4eqm1}) are manifestly invariant 
under $\l(t)\to\l(t) +\l_0$, where $\l_0$ is an arbitrary constant. 
  
It is straightforward to calculate how the right-hand-side of eq.~(\ref{4eqm}) 
transforms under the duality (\ref{duality}) by differentiating 
eq.~(\ref{duality}). We find
\begin{align}
\fracmm{\dt{a}}{a}~=~& -\fracmm{\dt{b}}{b}~~,\quad 
\fracmm{\ddt{a}}{a}= -\fracmm{\ddt{b}}{b} 
+2\left(\fracmm{\dt{b}{}}{b}\right)^2~~,\nonumber\\
\fracmm{\dddt{a}}{a}~=~& -\fracmm{\dddt{b}}{b} + 6\left(\fracmm{\dt{b}}{b}
\right)
\left(\fracmm{\ddt{b}}{b}\right)-6\left(\fracmm{\dt{b}}{b}\right)^3~,\\
\fracmm{\ddddt{a}}{a}~=~& -\fracmm{\ddddt{b}}{b} 
+ 6\left(\fracmm{\ddt{b}}{b}\right)^2
+8\left(\fracmm{\dt{b}}{b}\right)\left(\fracmm{\dddt{b}}{b}\right)-
36\left(\fracmm{\dt{b}}{b}\right)^2\left(\fracmm{\ddt{b}}{b}\right)
+24\left(\fracmm{\dt{b}}{b}\right)^4 \nonumber 
\lb{deri} 
\end{align}

To see how the duality affects the polynomial $P_8$, we consider the case with 
the 3rd order time derivatives, motivated by eq.~(\ref{br2frw}). 
We introduce the notation 
\be
\fracmm{\dt{a}}{a}=x~,\quad \fracmm{\ddt{a}}{a}=y~, \quad 
\fracmm{\dddt{a}}{a}=z \lb{tnot}
\ee
so that the duality invariance condition reads
\be
P_8(-x,2x^2-y,6xy-6x^3-z)=P_8(x,y,z) \lb{tcon}
\ee 
The structure of the polynomial $P_8$ in eq.~(\ref{poly}), as the sum over 
partitions of $8$, restricts a solution to eq.~(\ref{tcon}) to be most 
quadratic in $z$,
\be  
P_8(x,y,z)= a_2(x,y)z^2 +b_5(x,y)z + c_8(x,y) \lb{struc}
\ee
whose coefficients are polynomials in $(x,y)$, of the order being given by 
their subscripts, i.e.
\begin{align}
a_2(x,y)=~& a_0 x^2 +a_1y~,\nonumber\\
b_5(x,y)=~& b_0x^5 +b_1x^3y +b_2xy^2~,\\
c_8(x,y)=~& c_4y^4+c_3y^3x^2+c_2y^2x^4+c_1yx^6+c_0x^8\nonumber 
\lb{3polys}
\end{align} 
After a substitution of eqs.~(\ref{struc}) and (6.14) into eq.~(\ref{tcon}), 
we get an {\it over\/}determined system of linear equations 
on the coefficients. Nevertheless, we find that there is a consistent general 
solution,
\begin{align} 
P_8(x,y,z)= &~ a_0x^2z^2 +(b_0x^5-3a_0xy^2)z \nonumber \\
&~ +c_4y^4 +(9a_0-4c_4)y^3x^2+c_2y^23x^4  \\
&~ + (8c_4-18a_0-3b_0-2c_2)yx^6+c_0x^8 \nonumber 
\lb{answ}
\end{align}
parameterized by merely five real coefficients $(a_0,b_0,c_4,c_2,c_0)$. 
Requiring the existence of the exact solution (\ref{expf}), i.e. the 
positivity of $(\#)$ in eq.~(\ref{simple}), yields
\be 5c_4 + c_0 > 11a_0 +2b_0 + c_2 \lb{sumin} \ee

As regards the $(BR)^2$ gravity  representing the `minimal' candidate 
for the off-shell superstring effective action, 
we checked that neither the duality invariant structure (6.15)
 nor the inequality (\ref{sumin}) are satisfied by the coefficients present in 
eq.~(\ref{br2frw}). We interpret it as the clear indications that some 
additional Ricci-dependent terms {\it have to be added} to the $(BR)^2$ terms 
or, equivalently, the $(BR)^2$ gravity is ruled out as the off-shell effective 
action for superstrings.

Finally, we would like to mention about some possible simplifications and 
generalizations. 

The last equation (\ref{frwcurv}) apparently implies that the Ricci-dependent 
terms in $P_8$ should have the factor of $(y+2x^2)$. Hence, it may be possible 
to completely eliminate both the 4th and 3rd order time derivatives in our 
generalized Friedmann equations, though we are not sure that this choice is 
fully consistent. However, if so, instead of eq.~(\ref{tcon}) we would get
another duality condition, 
\be P_8(-x,2x^2-y)=P_8(x,y) \lb{anodual}
\ee
whose most general solution is simpler,
\be P_8(x,y)=c_0x^8 +c_5y(y-2x^2)\left[y(y-2x^2)-4x^6\right]
+c_6x^4y(y-2x^2)\lb{would}\ee
with merely three, yet to be determined coefficients $(c_0,c_5,c_6)$.

We would like to emphasize that our results above can be generalized to any 
finite order with respect to the spacetime curvatures in the off-shell 
superstring effective action, because it amounts to increasing the order of the
 polynomial $P$. The list (6.10) can be continued to any higher order in the 
derivatives. We can now speculate about the form of the generalized Friedmann 
equation to {\it all\/} orders in the curvature. It depends upon whether (i) 
there will be some finite maximal order of the time derivatives there, or 
(ii) the time derivatives of arbitrarily high order appear (we do not know 
about it). Given the case (i), we just drop the requirement that the 
right-hand-side of our cosmological equation (\ref{4eqm}) is a polynomial, and 
take a duality-invariant {\it function} $P$ instead. In the case (ii), we 
should replace the function by a {\it functional}, thus getting a non-local 
equation having the form
\be \lb{Aeq} H^2 = \fracmm{\dt{a}{}^2}{a^2}= \b P[a(t)] \ee
whose functional $P$ is subject to the non-trivial duality constraint
\be \lb{tdual} P[a(t)]=P[1/a(t)]~. \ee

Imposing {\it simultaneously} both conditions of stability and duality 
invariance leads to severe constraints on the $c$-coefficients. 
Hence, it also severely restricts the quantum ambiguities in the 
superstring-generated quartic curvature gravity. Finding their solutions seems
to be a non-trivial mathematical problem. We would like to investigate it 
elsewhere \cite{prog}.

\section{Conclusion}

The higher curvature terms in the gravitational action defy the famous 
Hawking-Penrose theorem \cite{hawp} about the existence of a spacetime 
singularity in any exact solution to the Einstein equations. As we demonstrated
 in this paper, the initial cosmological singularity can be easily avoided by 
condsidering the superstring-motivated higher curvature terms on 
equal footing (i.e. non-perturbatively) with the Einstein-Hilbert term.

Our results predict the possible existence of the very short de Sitter 
phase driven by the quartic curvature terms, in the early inflationary epoch.

Though we showed the natural existence of inflationary (de Sitter) exact 
solutions without a spacetime singularity under rather generic conditions on 
the coefficients in the higher-derivative terms, it is not enough for robust 
physical applications. As a matter of fact, we assumed the dominance
of the higher curvature gravitational terms over all matter contributions in 
the very early Universe at the Planck scale. However, given the expansion of 
the Universe under the geometrical inflation, the spacetime curvatures should 
decrease, so that the matter terms can no longer be ignored. The latter may 
effectively replace the geometrical inflation by another matter-dominated 
mechanism, thus allowing the inflation to continue substantially below the
 Planck scale. 

In addition, the number of e-foldings (\ref{efol}) is just about one in our 
scenario based on the quartic curvature terms, which makes it difficult to 
compete with the conventional inflation mechanisms \cite{inf}. An investigation
of the possible `Graceful Exit' strategies, towards a matter-driven inflation
 is, however, beyond the scope of the given paper. 

The quartic curvature terms are also relevant to the Brandenberger-Vafa 
cosmological scenario of string gas cosmology \cite{bvafa} --- see e.g. 
ref.~\cite{borunda} for a recent investigation of the higher curvature 
corrections there.~\footnote{The higher curvature terms  were considered only 
perturbatively in ref.~\cite{borunda}.} 

The higher time derivatives in the equations of motion may be unavoidable when
using the higher curvature terms, but we do not see that they constitute a 
trouble.

Gravity with the quartic curvature terms is a good playground for going beyond 
the Einstein equations. Our analysis may be part of a more general approach 
based on superstrings, including dynamical moduli and extra dimensions.
  
\section*{Acknowledgements}

One of authors (SVK) would like to thank the Institute for Theoretical
Physics, Leibniz University of Hannover, Germany, for kind hospitality extended
to him during part of this investigation. This work is partially supported by 
the Japanese Society for Promotion of Science (JSPS) under the Grant-in-Aid 
programme for scientific research, and the bilateral German-Japanese exchange
programme under the auspices of JSPS and DFG (Deutsche Forschungsgemeinschaft).

We are grateful to D. Berenstein, E. Elizalde, A. Hebecker, L. Kofman, 
O. Lechtenfeld, U. Lindstrom, K. Maeda, N. Ohta, N. Sakai, I. Shapiro and 
M. Vasquez-Mozo for discussions and correspondence. We also thank the referees
for their constructive remarks.

\newpage

\section{Appendix A: our notation, and identities}
\def\theequation{\thesection.\arabic{equation}}
\makeatother


\newcommand{\ga}{\alpha}
\newcommand{\gb}{\beta}
\newcommand{\gc}{\gamma}
\newcommand{\gd}{\delta}
\newcommand{\gz}{\zeta}
\newcommand{\gq}{\theta}
\newcommand{\gi}{\iota}
\newcommand{\gk}{\kappa}
\newcommand{\gl}{\lambda}
\newcommand{\gs}{\sigma}
\newcommand{\gu}{\upsilon}
\newcommand{\gw}{\omega}

\newcommand{\gA}{\Alpha}
\newcommand{\gB}{\Beta}
\newcommand{\gC}{\Gamma}
\newcommand{\gD}{\Delta}
\newcommand{\gZ}{\Zeta}
\newcommand{\gQ}{\Theta}
\newcommand{\gI}{\Iota}
\newcommand{\gK}{\Kappa}
\newcommand{\gL}{\Lambda}
\newcommand{\gS}{\Sigma}
\newcommand{\gU}{\Upsilon}
\newcommand{\gW}{\Omega}

\renewcommand{\(}{\left(}
\renewcommand{\)}{\right)}
\renewcommand{\[}{\left[}
\renewcommand{\]}{\right]}

We use the basic notation of ref.~\cite{ll} with the signature $(+,-,-,-)$. 
The (Riemann-Christoffel) curvature tensor is given by
\be
R\ud{i}{klm} = \fracmm{\pa\G^i_{km}}{\pa x^l}-\fracmm{\pa\G^i_{kl}}{\pa x^m} + 
\G^i_{nl}\G^n_{km} -
\G^i_{nm}\G^n_{kl} \lb{cur}
\ee
in terms of the Christoffel symbols
\be
\G ^i_{kl} = \ha g^{im}\left( \fracmm{\pa g_{mk}}{\pa x^l}
+\fracmm{\pa g_{ml}}{\pa x^k}
-\fracmm{\pa g_{kl}}{\pa x^m} 
\right)\lb{chs}
\ee
It follows
\be R_{iklm} =\ha\left( \fracmm{\pa^2 g_{im}}{\pa x^k\pa x^l}
+\fracmm{\pa^2 g_{kl}}{\pa 
x^i\pa x^m}
-\fracmm{\pa^2 g_{il}}{\pa x^k\pa x^m} -\fracmm{\pa^2 g_{km}}{\pa x^i\pa x^l}
\right)+g_{np}
\left(
 \G^n_{kl}\G^p_{im} - \G^n_{km}\G^p_{il}\right)\lb{cur2}
\ee
The traceless part of the curvature tensor is given by a Weyl tensor,
\be
C_{ijkl}= R_{ijkl} -\fracm{1}{2}\left( g_{ik}R_{jl}-g_{jk}R_{il}-g_{il}R_{jk}
+g_{jl}R_{ik}\right)+\fracm{1}{6}\left( g_{ik}g_{jl}- g_{jk}g_{il}\right)R
\lb{weyl}
\ee
where we have introduced the Ricci tensor and the scalar curvature,
\be R_{ik} =g^{lm}R_{limk}~,\quad R =g^{ik}R_{ik}\lb{rics}\ee
 
The dual curvature is defined by
\be {}^*R_{iklm}=\ha E_{ikpq}R\ud{pq}{lm}\lb{dcur}\ee
where $E_{iklm}=\sqrt{-g}\,\ve_{iklm}$ is Levi-Civita tensor.

The {\it Euler} (E)  and {\it Pontryagin} (P) topological densities in four 
dimensions are 
\be E_4= \fracm{1}{4}\ve_{ijkl}\ve^{mnpq}R\ud{ij}{mn}R\ud{kl}{pq}=
 {}^*R_{ijkl}{}^*R^{ijkl}
 \lb{eu}
\ee
and
\be P_4= {}^*R_{ijkl}R^{ijkl} \lb{pont}
\ee
respectively.

The {\it Bel-Robinson} (BR) tensor is defined by \cite{br}
\begin{align}
T_R^{iklm} & = R^{ipql}R\udu{k}{pq}{m} +{}^*R^{ipql}{}^*R\udu{k}{pq}{m} 
\nonumber\\
 & =   R^{ipql}R\udu{k}{pq}{m} +R^{ipqm}R\udu{k}{pq}{l}-\ha g^{ik}R^{pqrl}
R\du{pqr}{m}
\lb{bro}
\end{align}

Its Weyl cousin is given by
\be
T_C^{iklm}= C^{ipql}C\udu{k}{pq}{m} +C^{ipqm}C\udu{k}{pq}{l}
-\ha g^{ik}C^{pqrl}C\du{pqr}{m}
\lb{wbrob}
\ee

The Riemann-Christoffel curvature (modulo Ricci-dependent terms) is most 
easily described in the Petrov formalism \cite{petrov} by imposing the 
Ricci-flatness condition $R_{ik}=0$. A metric $g_{mn}$ at a given point in 
space-time can always be brought into Minkowski form $\h={\rm diag}(+,-,-,-)$, 
whereas the curvature tensor components can be represented by~\footnote{We 
use the lower-case greek letters to represent vector indices in three (flat) 
spatial dimensions.}
\begin{align}
	A_{\ga\gb}=R_{0\ga 0\gb},~~
	C_{\ga\gb}=\fr{1}{4}\epsilon_{\ga\gc\gd}\epsilon_{\gb\gl\mu}
R_{\gc\gd\gl\mu},~~
	B_{\ga\gb}=\fr{1}{2}\epsilon_{\ga\gc\gd}R_{0\gb\gc\gd}
\end{align}
where the 3d tensors $A$ and $C$ are symmetric by definition, 
$\ga,\gb,\cdots=1,2,3$, and 
$\epsilon_{\ga\gb\gc}$ is 3d Levi-Civita symbol normalized by 
$\epsilon_{123}=1$.

The Ricci-flatness condition
implies that $A$ is traceless, $B$ is symmetric, and $C=-A$. It is now natural 
to introduce a symmetric (traceless) complex 3d tensor  
\begin{align}
	D_{\ga\gb}=A_{\ga\gb}+iB_{\ga\gb}
\end{align}
and bring it into one of its canonical (Petrov) forms, called I, II or III, 
depending upon a number (3, 2 or 1, respectively) of eigenvectors of $D$. For 
our purposes, it is most convenient to use the form I with three independent 
(complex) eigenvectors, so that the real matrices $A$ and $B$ can be 
simultaneously diagonalized as
\begin{align}
	A_{\ga\gb}&=\mr{diag}(\ga',~\gb',~-\ga'-\gb')~, \nonumber \\
	B_{\ga\gb}&=\mr{diag}(\ga'',~\gb'',~-\ga''-\gb'') \lb{eigen}
\end{align}
in terms of their real eigenvalues.

It is straightforward to write down any Riemann-Christoffel curvature 
invariants as the polynomials of their eigenvalues (\ref{eigen}) in the 
Petrov I form. It is especially useful for establishing various identities 
modulo Ricci-dependent terms. 
We find~\footnote{We are grateful to A. Morishita for his help with 
calculations.}
\be R_{mnpq}R^{mnpq}=16\left( \ga'^2 + \gb'^2+\ga'\gb'-\ga''^2-\gb''^2
-\ga''\gb'' \right)
\lb{eigen1}
\ee
and
\be 
{}^*R_{mnpq}R^{mnpq}=16\left( -2\ga'\ga''-2\gb'\gb''-\ga''\gb'-\ga'\gb'' 
\right)\lb{eigen2}
\ee
so that
\begin{align}
	&\!\!\!\!\!\!\!\!\! (R_{mnpq}R^{mnpq})^2+({}^*R_{mnpq}R^{mnpq})^2 
\lb{eigen3} \\
	=2^8~(
		&\ga'^4+\gb'^4+\ga''^4+\gb''^4+2\ga'^2\ga''^2+2\gb'^2\gb''^2 
\nonumber \\
		&+2\ga'^3\gb'+2\ga'\gb'^3+2\ga''^3\gb''+2\ga''\gb''^3 
\nonumber\\
		&+2\ga'^2\ga''\gb''+2\ga''\gb'^2\gb''+2\ga'\ga''^2\gb'
+2\ga'\gb'\gb''^2
\nonumber \\
	 &+3\ga'^2\gb'^2+3\ga''^2\gb''^2-\ga'^2\gb''^2-\ga''^2\gb'^2
+8\ga'\ga''\gb'\gb'')
\nonumber
\end{align}
Similarly one finds
\be 
{}^*R_{mnpq}{}^*R^{mnpq}=-16\left( \ga'^2 + \gb'^2+\ga'\gb'-\ga''^2-\gb''^2
-\ga''\gb'' \right)
\lb{eigen4}
\ee
For example, when being compared to eq.~({\ref{eigen1}), it yields the identity
\be {}^*R_{mnpq}{}^*R^{mnpq}=-R_{mnpq}R^{mnpq} +{\cal O}(R_{mn})\lb{iden1}\ee

As regards the superstring correction (\ref{oco4}) in four dimensions, we find
\begin{align}
J_R=8(&\ga'^4+\gb'^4+\ga''^4+\gb''^4+2\ga'^2\ga''^2+2\gb'^2\gb''^2 \nonumber \\
		&+2\ga'^3\gb'+2\ga'\gb'^3+2\ga''^3\gb''+2\ga''\gb''^3 
\nonumber \\
		&+2\ga'^2\ga''\gb''+2\ga''\gb'^2\gb''+2\ga'\ga''^2\gb'
+2\ga'\gb'\gb''^2 
\nonumber \\
		&+3\ga'^2\gb'^2+3\ga''^2\gb''^2-\ga'^2\gb''^2-\ga''^2\gb'^2
+8\ga'\ga''\gb'
\gb'')
\lb{eigen5}
\end{align}
The BR tensor (\ref{bro}) squared in the Petrov I form reads
\begin{align}
T_{mnpq}T^{mnpq} &~~ = 2^6~( \ga'^4+\gb'^4+\ga''^4+\gb''^4+2\ga'^2\ga''^2
+2\gb'^2\gb''^2 
\lb{eigen6} \\
	&~~+2\ga'^3\gb'+2\ga'\gb'^3+2\ga''^3\gb''+2\ga''\gb''^3 \nonumber \\
	&~~+2\ga'^2\ga''\gb''+2\ga''\gb'^2\gb''+2\ga'\ga''^2\gb'
+2\ga'\gb'\gb''^2 \nonumber \\
	&~~+3\ga'^2\gb'^2+3\ga''^2\gb''^2-\ga'^2\gb''^2-\ga''^2\gb'^2
+8\ga'\ga''\gb'\gb'')
\nonumber
\end{align}
As a result, we find the identities
\be
T_{mnpq}T^{mnpq}=8J_R=\fr{1}{4}\[(R_{mnpq}R^{mnpq})^2+(R_{mnpq}^*R^{mnpq})^2\]
\lb{imp}
\ee	
which are valid on-shell, i.e. modulo Ricci-tensor-dependent terms.

\section{Appendix B: two-component formalism}

To complete our notation, we summarize basic definitions and main features of 
the two-component spinor formalism in gravitation ({\it cf}. 
refs.~\cite{wald,wb}). The main point is the use of an $sl(2;{\bf C})$ algebra 
isomorphic to the Lorentz algebra $so(1,3;{\bf R})$.

We use lower-case (middle) latin indices for the curved space-time vector 
indices, capital (early) latin letters for the tangent (flat spacetime) vector 
indices, and lower-case (early) greek letters for the (tangent spacetime) 
spinor indices, $i,j,k,\ldots=0,1,2,3$ and $A,B,C,\ldots=0,1,2,3$, whereas 
$\a,\b,\ldots=1,2$ and $\dt{\a},\dt{\b},\ldots=\dt{1},\dt{2}$.

A four-component Dirac spinor $\J$ can be decomposed into its chiral and 
anti-chiral parts, $\j_{\a}$ and $\bar{\j}^{\dt{\b}}$,  by using the chiral 
projectors $\G_{\pm}=\ha(1\pm \g_5)$, 
where $\g^2_5=1$. The simplest form of chiral decomposition is obtained in the 
basis for Dirac 
gamma matrices with a diagonal $\g_5=i\g_0\g_1\g_2\g_3$ matrix,
\be \g^A= \left( \begin{array}{cc} 0 & \s^A_{\a\dt{\b}} \\ 
\tilde{\s}^{A\dt{\b}\a} & 0\end{array}\right)
~,\qquad \s^A=({\bf 1},i\vec{\s}),\quad \tilde{\s}^A=({\bf 1},-i\vec{\s}) 
\lb{gdec} \ee 
Here ${\bf 1}$ is a unit $2\times 2$ matrix, and $\vec{\s}$ are three Pauli 
matrices.
  
Given a vector field $V_i(x)$ in a curved spacetime, it can always be 
represented by a bispinor field $V_{\a\dt{\b}}(x)$,
\be V_{\a\dt{\b}} = V_ie^i_A \s^A_{\a\dt{\b}}~~,\quad
V_i = e_i^B  \frac{1}{2}V_{\a\dt{\b}}\tilde{\s}_B^{\dt{\b}\a}
 \lb{vcon}\ee
where we have introduced the vierbein $e_A^i(x)$, together with its inverse 
$e^A_i(x)$, obeying the relations
\be   g_{ij}e^i_Ae^j_B=\h_{AB}~,\quad \h_{AB}e^A_ie^B_j=g_{ij}\lb{vierb}\ee

For instance, one easily finds that the metric in the two-component formalism 
can be represented by a product of two Levei-Civita symbols,
\be
 g_{\a\b\dt{\a}\dt{\b}}=\ve\low{\a\b}\ve_{\dt{\a}\dt{\b}} \lb{twom}
\ee
    
As regards the curvature tensor, it can be naturally decomposed in the 
two-component formalism as follows:
\begin{align}
 R_{\a\b\g\d\dt{\a}\dt{\b}\dt{\g}\dt{\d}} = &  
~~C\low{\a\b\g\d}\ve_{\dt{\a}\dt{\b}}
\ve_{\dt{\g}\dt{\d}} + 
\Bar{C}_{\dt{\a}\dt{\b}\dt{\g}\dt{\d}}\ve\low{\a\b}\ve\low{\g\d}  \\
&   + D_{\dt{\a}\dt{\b}\g\d}\ve\low{\a\b}\ve_{\dt{\g}\dt{\d}} + 
\Bar{D}_{\a\b\dt{\g}\dt{\d}}\ve_{\dt{\a}\dt{\b}}\ve\low{\g\d} \nonumber \\
& + E\left( \ve\low{\a\g}\ve\low{\b\d}+ \ve\low{\a\d}\ve\low{\b\g}\right) 
\ve_{\dt{\a}\dt{\b}}
\ve_{\dt{\g}\dt{\d}} +
  \Bar{E}\left( \ve_{\dt{\a}\dt{\g}}\ve_{\dt{\b}\dt{\d}}+ 
\ve_{\dt{\a}\dt{\d}}\ve_{\dt{\b}\dt{\g}}
 \right)\ve\low{\a\b}\ve\low{\g\d}\nonumber \lb{cur2f}
\end{align}
The four-spinor $C$ (or $\bar{C}$) is {\it totally symmetric} with respect to 
its chiral (or anti-chiral) spinor indices, while is is also traceless, thus 
representing the self-dual (or anti-self-dual) part of the Weyl tensor 
(\ref{weyl}),
\be 
C_{\a\b\g\d\dt{\a}\dt{\b}\dt{\g}\dt{\d}} = C\low{\a\b\g\d}\ve_{\dt{\a}\dt{\b}}
\ve_{\dt{\g}\dt{\d}}+\Bar{C}_{\dt{\a}\dt{\b}\dt{\g}\dt{\d}}\ve\low{\a\b}
\ve\low{\g\d}
\lb{2fweyl}\ee  
The four-spinor $D_{\dt{a}\dt{\b}\g\d}$ is symmetric with respect to its first 
two indices, as well as with respect to the last two indices, while it is also 
traceless, thus representing  the traceless part of the Ricci tensor, 
\be 
R_{\a\g\dt{\a}\dt{\g}}= \ve^{\d\b}\ve^{\dt{\d}\dt{\b}}
R_{\a\b\g\d\dt{\a}\dt{\b}\dt{\g}\dt{\d}}
\lb{2fricci}
\ee
The scalar $E=\Bar{E}$ represents the scalar curvature $R$.
One easily finds
\be
R_{\a\b\dt{\a}\dt{\b}}=-2D_{\dt{\a}\dt{\b}\a\b}+6E\ve_{\dt{\a}\dt{\b}}
\ve\low{\a\b}~,\qquad R=24E \lb{2fsca}\ee

The Bianchi II identities $\nabla_{[m}R_{ij]kl}=0$ in the two-component 
formalism read as follows:
\be 
\nabla\ud{\a}{\dt{\b}}C\low{\a\b\g\d}= 
\nabla\low{(\b}{}^{\dt{\a}}D_{\g\d)\dt{\a}\dt{\b}}~~,
\qquad
\nabla^{\g\dt{\a}}D_{\g\d\dt{\a}\dt{\b}}+ 3\nabla_{\d\dt{\b}}E=0 \lb{2fb2}
\ee

\end{document}